\documentclass[prl]{revtex4}
\def\r{\hat\rho}
\def\P{\hat P}
\def\G{\Gamma}
\def\L{{\cal L}}
\def\s{\sigma}
\def\p{\hat p}\def\x{\hat x}
\def\a{\alpha}
\def\xb{\bar   x}\def\pb{\bar   p}
\def\xt{\tilde x}\def\pt{\tilde p}\def\Gt{\tilde\G}

\def\ket#1{|#1\rangle}
\def\bra#1{\langle#1|}

\def\be{\begin{equation}}
\def\ee{\end{equation}}
\def\bea{\begin{eqnarray}}
\def\eea{\end{eqnarray}}
\def\lb{\label}

\newcommand{\I}{\mbox{i}}
\newcommand{\D}{\mbox{d}}

\begin{document}
\title{Robustness and diffusion of pointer states}
\author{Lajos Di\'osi}
\affiliation
{Research Institute for Particle and Nuclear Physics,
P.O. Box 49, H-1525 Budapest 114, Hungary.}
\author{Claus Kiefer}
\affiliation
{Fakult\"at f\"ur Physik, Universit\"at Freiburg,
Hermann-Herder-Stra\ss e 3, D-79104 Freiburg, Germany.}
\date{\today}
\begin{abstract}
Classical properties of an open quantum system emerge through its 
interaction with other degrees of freedom (decoherence). 
We treat the case where this interaction produces a Markovian 
master equation for the system. We derive the corresponding 
distinguished local basis (pointer basis) by three methods.
The first demands that the pointer states mimic as close as 
possible the local non-unitary evolution. The second demands that 
the local entropy production be minimal (predictability sieve). 
The third imposes robustness on the inherent quantum and emerging 
classical uncertainties. All three methods lead to localized 
Gaussian pointer states,
their formation and diffusion being governed by 
well-defined quantum Langevin equations.
\end{abstract}
\maketitle
The programme of decoherence has been very successful in explaining
the classical appearance of macroscopic or mesoscopic quantum systems,
both theoretically and experimentally \cite{deco}. The interaction
of a quantum system with its environment leads in these cases
to a delocalization of phase relations in the full configuration
space of system plus environment, preventing them to be observed
locally, i.e. at the system itself. Thereby the environment
distinguishes a certain preferred basis for the system, which
can be used to describe the apparent classical behavior
({\em pointer basis} 
\cite{Zur81}). An important question
is how the pointer basis can be determined and whether it is
{\it uniquely} fixed. It would then be possible to give a unique 
decomposition of the reduced density matrix into an
apparent ensemble of wave functions. 

No unique rules have so far been adopted to
calculate the pointer states in the general case. 
In \cite{Zeh73} the suggestion was made that the pointer
basis (there called
collection of ``memory states'') is characterized
by its {\it robustness} (there called ``dynamical stability''). 
A first quantitative measure
to investigate the dynamical stability is the rate of
de-separation introduced in \cite{Kue73} -- it measures
how fast a quantum system becomes entangled with environmental
degrees of freedom. In a model consisting of harmonic
oscillators, it was shown that coherent states are the most
stable states and therefore can be considered as pointer states
\cite{Kue73,deco}. 
A different measure for robustness
is the ``predictability sieve'' put forward in \cite{Zur93}.
The pointer basis is there distinguished by the property
of having the least production rate
for local entropy during the coupling to the environment. In the case of
harmonic oscillators, this again leads to the coherent states
as the pointer basis \cite{ZurHabP93}. At least for such simple systems,
rate of de-separation and predictability sieve are roughly
equivalent measures \cite{deco}.

On the other hand, the theory and formalism of quantum state diffusion
(QSD) were put forward to attribute
random wave functions for local systems,
which satisfy an appropriate Langevin equation \cite{qsd}. 
These wave functions are known to be related to possible continuous 
measurements \cite{Dio88a}
as well as to decoherent histories \cite{DioGisHP95}
of the given local
system. But even if we take them as wave functions of mere formal
meaning (since subsystems do not, in general,
possess their own pure states), the question arises whether
there is any connection between these states and the pointer
states, in cases where the local system exhibits
classical properties. We shall
show that there is in fact such a connection -- 
pointer basis and QSD basis are substantially the same.
For this aim, we shall also present below a new, alternative,
derivation of QSD.
    
In the following we shall
consider the dynamics of the reduced density matrix,
$\r(t)$, of a system interacting with a certain
decohering environment. Ideal {\it pointer states}
(described by a fixed set of projectors $\P_n$) 
are characterized by the fact that $\r(t)$ can be decomposed as
\be\lb{rho0}
\r(t)\rightarrow\sum_n f_n\P_n\ ,~~~~~t\gg t_D \ , 
\ee
for a generic initial state $\r(0)$, where $t_D$ is the decoherence
time. 
The weights $\{f_n\}$ correspond to a normalized probability 
distribution. The pointer states 
$\{\P_n;n=1,2,\dots\}$ form in this case an orthogonal system.
For macroscopic systems, $t_D$ is extremely short
\cite{deco,JooZeh85}.
More generally, one would expect
\be
\r(t)\rightarrow\int f(\G)\P(\G)\D\G\ ,~~~~~t\gg t_D \ , \lb{rho}
\ee
where $f(\G)$ is a probability distribution over the
pointer states $\P(\G)$ which project now on
an overcomplete set of pure states (the above-mentioned coherent
states provide an example for this). The pointer states in
(\ref{rho0}) or (\ref{rho}) result after an explicit interaction with the
environment is taken into account \cite{deco}.

In many cases, the effects of decoherence can be described
by the following Markovian master equation \cite{deco},
\bea
\frac{\D\r(t)}{\D t}\equiv\L\r(t)
                 &=&-\frac{\I}{2m}[\p^2,\r(t)]-
               \frac{D}{2}[\x,[\x,\r(t)]]\ ,
 \lb{master}
\eea
where $D$ describes the strength of the interaction with the
environment. Such an equation arises, for example,
in situations where environmental degrees of freedom scatter
off a macroscopic object and localize it
by carrying away quantum correlations with
the object \cite{JooZeh85,deco}.
Applying the concept of predictability sieve would mean to minimize
the local production of ``linear entropy''
 $S(t)=1-\mbox{tr}\r^2(t)$.
This does not give a unique answer,
since the result depends explicitly on $t$. In the oscillator case,
one has therefore calculated the time-integrated entropy production
\cite{ZurHabP93,deco}. If one considers
 the initial entropy production
rate $\dot S(0)$,
assuming the initial state $\r(0)$ of the subsystem to be a pure
state $\P(\G)$, one finds from (\ref{master}) that 
$
\dot S(0)  
  = D\left(\langle\x^2\rangle-\langle\x\rangle^2\right)
 \equiv D\s^2\ ,
$
where the expectation value refers to the initial state.
This rate would be minimized
if the pointer wave functions were delta functions. However, 
their spread $\s$ increases dynamically due to the kinetic term
in (\ref{master}). Therefore, very narrow wave functions
do not produce minimum entropy on a finite time scale and thus cannot
correspond to pointer states.
The (coherent) unitary spreading and the (incoherent) non-unitary 
localizing terms of the master
equation (\ref{master}) are competing with each other.
For a wave function of characteristic width $\s$,
the above two effects 
are approximately 
balanced for the ``equilibrium width" \cite{JooZeh85,Dio87a,deco}:
\be
\s_0 \sim \left(Dm\right)^{-1/4}\ . \lb{width}
\ee 
It is then reasonable to conjecture that,
in the spirit of predictability sieve and of dynamical
robustness, $\s_0$ will be 
the characteristic width of the pointer states. This is in fact what
we shall show by using three different methods, all invoking
a principle of robustness.

The first method goes as follows.
Let us allow for the pointer 
state $\P(\G)$ a certain natural time dependence such that 
it may initially evolve as close as possible
along the true state $\r$ satisfying the master equation (\ref{master}),
and then reach a stationary state.
We introduce the ``speed'' $v$ describing the departure of the states
$\P(\G)$ from $\r$ in the Hilbert-Schmidt norm:
\be
v^2=\mbox{tr}\left[\frac{\D}{\D t}\P(\G)-\L\P(\G)\right]^2\ . \label{HS} 
\ee
The smaller $v$, the greater is the robustness of the pointer
states $\P(\G)$. Hence one defines the optimum drift of the
{\it pure} pointer state $\P(\G)\equiv\psi\psi^\dagger$
by minimizing $v$ \cite{GisRig95}. 
This is given by the nonlinear
equation \cite{Dio86,GisRig95}:
\be
\dot\psi=(\L\psi\psi^\dagger)\psi-\langle\L\psi\psi^\dagger\rangle\psi\ .
\lb{drift}
\ee
This result is valid for all kinds of
 Markovian subdynamics. In our special case
(\ref{master}), it yields the following nonlinear wave equation:
\be
\dot\psi=-\frac{\I}{2m}\p^2\psi
         -\frac{D}{2}\left[(\x-<\x>)^2-\s^2\right]\psi\ . \lb{driftpsi}
\ee
As shown in~\cite{Dio87b},
this equation has a stationary solution which is {\it unique} up to
Galilean transformations. The wave function of the fiducial 
stationary state is the complex Gaussian wave packet
\be
\psi_0(x)=\left(\a_R/2\pi\right)^{1/4}\exp(-\a x^2/4)\ , \lb{gauss0} 
\ee
with parameter
\be\lb{aHS}
\a\equiv\a_R+\I\a_I=(1-\I)\sqrt{2Dm}\ .
\ee
The principle of ``Hilbert-Schmidt robustness'' has thus
singled out unique Gaussian 
pointer states as the robust pure states closest to the true 
non-unitary local dynamics. The exact width $\s\equiv 1/\sqrt{\a_R}$ confirms 
the heuristic estimate (\ref{width}). 
Accordingly, we restrict our further discussion to pointer 
states $\P(\G)$ with Gaussian wave functions and make the ansatz
\be
\psi_{\G}(x)=\left(\a_R/2\pi\right)^{1/4}
  \exp\left(-\a (x-\xb)^2/4+\I\pb
   (x-\xb)\right)\ , \lb{gauss} 
\ee
where $\G\equiv(\xb,\pb)^T$ has been understood. 
For later purposes, we calculate the correlation matrix ${\bf C}$   
\be\lb{C}
{\bf C}\equiv\bra{\psi_0}
\left(\begin{array}{cc}\hat x^2&(\hat x\hat p+h.c.)/2\\
(\hat x\hat p+h.c.)/2&\hat p^2\end{array}\right)\ket{\psi_0}
=\frac{1}{\a_R}\left(\begin{array}{cc}1&-\a_I/2\\
                   -\a_I/2&\vert\a\vert^2/4\end{array}\right)
\ee
of the quantum uncertainties of the 
canonical observables in the pointer state itself,
where $\psi_0$ denotes the fiducial state (\ref{gauss0}).
It was shown in \cite{JooZeh85} that the states diagonalizing
$\r$ exactly are the harmonic oscillator eigenfunctions
which are very broad, while narrow eigenfunctions
are apparently obtained only for discrete systems.
In contrast to these, the above pointer
states are well localized. For example, in the situation of
a small dust particle ($m=10^{-14}\ \mbox{g}$) scattered
by air molecules one has $D\sim 10^{32}\ \mbox{cm}^{-2}
\mbox{s}^{-1}$ \cite{JooZeh85} and therefore
$\s_0\approx(Dm)^{-1/4}\approx 10^{-11}\ \mbox{cm}$ and
$t_D\approx\sqrt{m/D}\approx 10^{-10}\ \mbox{s}$.

We now come to the second method. As a preparation,
we shall discuss the reduced dynamics of the local system in the
basis given by (\ref{gauss}).
We allow temporarily the parameter $\a$ to take an arbitrary 
complex value, and then derive again
a distinguished value.
If one allows a ``natural'' time dependence for the 
probability distribution $f(\G;t)$ of the pointer, 
the asymptotic condition (\ref{rho}) can be turned into an
exact identity:
\be
\r(t)=\int f(\G;t)\P(\G)\D\G,\quad
                                  t > t_D\ , \lb{rho1}
\ee
where $\D\G\equiv \D\xb \D\pb/2\pi$. This important fact will be
proven elsewhere \cite{DioKie00}. 
It generalizes the corresponding statement made in \cite{Dio87b}
for the specific value (\ref{aHS}) of $\a$
as well as the asymptotic statement proved in \cite{HalZou95,HalZou97}. 

 {}From (\ref{master}) and (\ref{rho1}) one can derive
an evolution equation for $f(\G;t)$:
\be
\frac{\D f(\G;t)}{\D t}=-\frac{\pb}{m}\partial_{\xb} f(\G;t)
        +\frac{1}{2}\left[ D_{pp}\partial^2_{\pb\pb}
                         +D_{xx}\partial^2_{\xb\xb}
                        +2D_{px}\partial^2_{\pb\xb} \right]f(\G;t)
\ , \lb{FP}
\ee
where the elements of the diffusion matrix are given by
\be
{\bf D}\equiv
\left(\begin{array}{cc}D_{xx}&D_{xp}\\D_{px}&D_{pp}\end{array}\right)
=\left(\begin{array}{cc}
-\a_I/m\a_R &\vert\a\vert^2/4m\a_R\\
\vert\a\vert^2/4m\a_R & D\end{array}\right)\ . \lb{D}
\ee
To find a formal solution of~(\ref{FP}), we use the Fourier 
representation
$
\tilde f(\Gt;t)
=\int f(\G;t)\exp\left[\I(\xt\pb-\pt\xb)\right]\D\G
$
with $\Gt=(\xt,\pt)^T$.
Eq.~(\ref{FP}) then leads to 
\be
\frac{\D {\tilde f}(\Gt;t)}{\D t}=
 -\frac{\pt}{m}\partial_{\xt}{\tilde f}(\Gt;t)
        -\frac{1}{2}\vert{\bf D}\vert\left[\Gt^T{\bf D}^{-1}
       \Gt\right]
        {\tilde f}(\Gt;t)\ , \lb{FPFou}
\ee
where $\vert{\bf D}\vert$ denotes the determinant of ${\bf D}$. 
The solution takes the form
\be
{\tilde f}(\Gt;t)
=\exp\left[-\frac{t}{2}\Gt^T{\bf G}(t)\Gt\right]
\tilde f(\xt-\pt t/m,\pt;0)\ . \lb{ftsol}
\ee 
By substitution into (\ref{FPFou}) one obtains explicitly the matrix of 
time-dependent coefficients:
\be
{\bf G}(t)=\left(\begin{array}{cc}D_{pp}&-D_{xp}+D_{pp}t/2m\\
            -D_{xp}+D_{pp}t/2m&D_{xx}-D_{xp}t/m+D_{pp}t^2/3m^2
               \end{array}\right)\ . \lb{Gt}
\ee

Eq.~(\ref{FP}) can be interpreted as a Fokker-Planck equation provided the
diffusion matrix ${\bf D}$ is non-negative. Then the weight function 
$f(\xb,\pb;t)$ of the pointer states $\P(\xb,\pb)$ will drift according to
the free-particle dynamics. At the same time 
{\em the state of the system will diffuse over the pointer 
states} $\P(\xb,\pb)$.
We can now implement the predictability sieve and minimize the production 
rate for linear entropy by minimizing the width of the Gaussian 
pointer states. In other words, we maximize $\a_R$ under the condition that 
the diffusion matrix be non-negative.
The condition that ${\bf D}$ has a non-negative determinant 
leads to the condition 
$
\a_R^4+2\a_R^2\a_I^2+
16Dm\a_R\a_I+\a_I^4\leq0\ 
$
which can only be fulfilled if $\a_I<0$, since $\a_R>0$ for (\ref{gauss}) 
to be normalizable. Introducing dimensionless polar coordinates $R,\phi$ by 
$\a\equiv\sqrt{Dm}R\exp(\I\phi)$, this condition reads
$
R^2+8\sin2\phi\leq0\ .
$
The maximum for $\a_R=\sqrt{Dm}R\cos\phi$ is reached if the
equality sign holds, since one could
otherwise increase $R$ by holding $\phi$ fixed and thus
increase $\a_R$. Maximizing $\a_R$ under the condition
$R^2=-8\sin2\phi$ then yields for $\a$ the following value
$\a_s$ distinguished by the predictability sieve:
\be 
\a_s= 3^{1/4}(\sqrt{3}-\I)\sqrt{Dm}\ , \lb{alphas}
\ee
which coincides, up to a small deviation in the
numerical coefficients, with the value 
(\ref{aHS}) following from the criterion of Hilbert-Schmidt robustness.  
This above slight departure of $\a_s$ might be related to the fact that
the given form of predictability sieve predicts a degenerate diffusion
matrix ${\bf D}$. We think, however, that the emerging incoherent 
uncertainties due to the pointer state diffusion must be made proportional
to the quantum uncertainties already present in the pointer states itself.
The ``robustness of uncertainties'' demands
that the matrix ${\bf C}$ (\ref{C}) of quantum correlations 
{\em be proportional} to the diffusion matrix
${\bf D}$ (\ref{D}) of the corresponding classical coordinates
for the pointer.
 From the condition that ${\bf C}=const\times{\bf D}$
we then obtain again the standard value (\ref{aHS}) for $\a$,
while ${\bf C}=m/2D\times{\bf D}$.

We shall now discuss our last method to determine the pointer
basis, which will involve quantum state diffusion.
As we see from (\ref{rho1}) and (\ref{FP}), the quantum state of the system,
when expanded as a mixture of pointer states, performes diffusion after the 
decoherence time has elapsed. This diffusion will, by construction, preserve
the shape (\ref{gauss0}) of the Gaussian wave packet, and only its center 
will walk randomly. It is then natural to ask, whether there is a
generic QSD process which, first, applies
 to {\em generic} pure initial states
and, second, tends to the above specific diffusion process for $t\gg t_D$.

As is well known, the Fokker-Planck equation (\ref{FP}) is equivalent to the 
 It\^{o}-Langevin equation
\cite{Ar}
\be\lb{Ito}
\D\G=V\D t + \D X\ ,
\ee
where $V=(\pb/m,0)$, and $\D X=(\D\xi,\D\pi)$ is the increment of a zero-mean 
Gaussian white noise with correlation matrix ${\bf D}\D t$. 
In case of phase-space diffusion the use of the It\^{o}-Langevin formalism 
instead of the Fokker-Planck formalism is a matter of taste. But the diffusion 
of the corresponding pointer states $\psi_{\G}$ would be quite awkward in 
the Fokker-Planck formalism.
We thus choose the It\^{o}-formalism 
and apply (\ref{Ito}) to the Gaussian pointer states (\ref{gauss}).
This leads to, substituting
 $\langle\x\rangle=\xb$
and $\langle\p\rangle=\pb$, 
\be\lb{qsdpsi}
\D\psi=-\frac{\I}{2m}\hat p^2\psi \D t 
  -\frac{D}{2}\left(\hat x -\langle\hat x\rangle\right)^2 \psi \D t 
      +(\x-\langle\x\rangle)\psi \D z~,
\ee
where the index $\G$ has been skipped.
The deterministic part of the evolution is governed, up to normalization,
by the same nonlinear wave equation (\ref{driftpsi}) which we had obtained
from the Hilbert-Schmidt robustness, 
while the random part is driven by the complex Gaussian white noise
\be\lb{dz}
\D z=\frac{\a}{2}\D\xi+\I\D\pi~.
\ee
Since the correlation matrix of
 $\D X=(\D\xi,\D\pi)$ is ${\bf D}\D t$, 
(\ref{D},\ref{dz}) yields  
\be\lb{dzdzstar}
M[\D z\D z^\star]=D\D t 
\ee
for the mean of the Hermitian correlation,
 independent of $\a$. But the 
correlation $M[\D z\D z]$ still depends on $\a$.
A most remarkable feature of (\ref{qsdpsi})
is that any reference to
the phase-space variables $\G=(\xb,\pb)$ has been cancelled.
For this reason we have
omitted the subscript $\G$ from $\psi$ and
extend the validity of the equation to
{\em arbitrary} initial state 
vectors. It is possible to prove \cite{Dio88b,GatGis91,HalZou95,Kol95}
that, starting from whatever
initial state $\psi(0)$,
the random solution $\psi(t)$
will tend to be the Gaussian pointer state
$\psi_{\G(t)}$, where $\G(t)$ is governed
by the diffusion process (\ref{Ito}). 
Eq.~(\ref{qsdpsi}) is called the
It\^{o}-Schr\"odinger equation of QSD.

Since the free 
parameter $\a$ still appears in $M[\D z\D z]$,
we are left with the non-uniqueness problem
of the QSD equations. If one, however, chooses the
distinguished value (\ref{aHS})
of $\a$, 
one finds, using~(\ref{dz}), the simple result  
\be\lb{dzdz}
M[\D z\D z]=M[\D z^\star \D z^\star]=0~, \ee
distinguishing a unique QSD.
Historically, this unique QSD was in the Fokker-Planck formalism 
singled out by certain invariance considerations 
\cite{Dio88c,Per90}.
The It\^o-Schr\"odinger equation (\ref{qsdpsi})
with the complex 
Gaussian white noise (\ref{dzdzstar},\ref{dzdz})
has become the dominating formalism of {\it standard} QSD 
theory \cite{qsd} extended for arbitrary Markovian reduced dynamics.
Applying exact forms of robustness criteria we have thus obtained a 
unique QSD which leads to stationary Gaussian pointer states
for $t\gg t_D$, whose centers undergo a diffusion process. 
With heuristic forms of robustness 
one could have chosen other QSD equations
like in ~\cite{Dio88b} ($\a_R=2\sqrt{Dm}$)
or \cite{HalZou95} ($\a_R=\sqrt{Dm}/2\sqrt{2}$). 
The recent proposal of ``maximal survival
probability'' from \cite{WV} differs from our first method
and does not lead to (\ref{dzdz}) \cite{DioKie00}.

In conclusion, we have demonstrated that three different methods of
dynamical robustness lead to an essentially {\it unique} local pointer
basis in case of Markovian local dynamics. The corresponding pointer 
states follow the classical trajectories up to a tiny random diffusion. 
Well-defined stochastic differential equations, known from the theory
of quantum state diffusion, govern both the formation and the diffusion 
of pointer states.   
These states can thus be used to characterize local
quasiclassical properties. The pointer states
are not an absolute property of the system in itself, but only
characterize certain stability properties with respect to 
interactions with the environment: They are least sensitive
to quantum entanglement, which is why interference terms
between them cannot be noticed by local observers.
They possess thus meaning with respect to an
observer-related branch of the total wave function or a
component corresponding to a potential fundamental
collapse \cite{deco,Zeh99}, while the
interaction with the environment is encoded in the choice
of our master equation (\ref{master}). 

We thank the Institute for
Advanced Study Berlin and the Isaac Newton Institute,
Cambridge, for their kind hospitality while this work
was begun. L.D. thanks the ESF QIT program and the Hungarian OTKA 
Grant 032640 for financial support. Critical comments by E.~Joos,
H.~Wiseman, and
H.D.~Zeh are gratefully acknowledged.


\begin{thebibliography}{99}
 \bibitem{deco} D. Giulini, E. Joos, C. Kiefer, J. Kupsch,
 I.-O. Stamatescu, and H.D. Zeh, {\em Decoherence and the Appearance
 of a Classical World in Quantum Theory} (Springer, Berlin, 1996). 
\bibitem{Zur81}  W.H. Zurek, Phys. Rev.~D {\bf 24}, 1516 (1981).
\bibitem{Zeh73} H.D. Zeh, Found. Phys. {\bf 3}, 109 (1973).
\bibitem{Kue73} O. K\"ubler and H.D. Zeh, Ann. Phys. (N.Y.)
        {\bf 76}, 405 (1973).
\bibitem{Zur93} W.H. Zurek, Progr. Theor. Phys. {\bf 89}, 281 (1993).
\bibitem{ZurHabP93} W.H. Zurek, S.~Habib, and J.P.~Paz,
        Phys. Rev. Lett.~{\bf 70}, 1187 (1993).
\bibitem{qsd} N. Gisin and I.C. Percival,
        J. Phys. A {\bf 25}, 5677 (1992);
        {\it ibid.\/} A {\bf 26}, 2233, 2245 (1993);
        I.C. Percival, {\em Quantum state diffusion}
        (Cambridge University Press, Cambridge, 1998).
\bibitem{Dio88a} L. Di\'osi, Phys. Lett. {\bf 129 A}, 419 (1988).
\bibitem{DioGisHP95} L. Di\'osi, N.Gisin, J.J. Halliwell,
                  and I.C. Percival,
        Phys. Rev. Lett. {\bf 74}, 203 (1995).
\bibitem{JooZeh85} E. Joos and H.D. Zeh,
                    Z. Phys. B {\bf 59}, 223 (1985). 
\bibitem{Dio87a} L. Di\'osi, Phys. Lett. {\bf 120 A}, 377 (1987).
\bibitem{GisRig95} N. Gisin and M. Rigo, J. Phys.~A
                    {\bf 28}, 7375 (1995).
\bibitem{Dio86} L. Di\'osi, Phys. Lett. {\bf 114 A}, 451 (1986).
\bibitem{Dio87b} L. Di\'osi, Phys. Lett. {\bf 122 A}, 221 (1987).
\bibitem{DioKie00} L. Di\'osi and C. Kiefer, in preparation.
\bibitem{HalZou95} J.J. Halliwell and A. Zoupas,
        Phys. Rev. A {\bf 52}, 7294 (1995).
\bibitem{HalZou97} J.J. Halliwell and A. Zoupas,
        Phys. Rev. A {\bf 55}, 4697 (1997).
\bibitem{Ar} L. Arnold, {\em Stochastic differential equations: Theory and
    application} (Wiley, New York, 1974).
 \bibitem{Dio88b} L. Di\'osi, Phys. Lett. {\bf 132 A}, 233 (1988).
\bibitem{GatGis91} D. Gatarek and N. Gisin,
         J. Math. Phys. {\bf 32}, 2152 (1991).
\bibitem{Kol95} V.N. Kolokoltsov,
         J. Math. Phys. {\bf 36}, 2741 (1995).
\bibitem{Dio88c} L. Di\'osi, J. Phys. A {\bf 21}, 2885 (1988).
\bibitem{Per90} I.C. Percival, 
        London University reports QMW DYN 90-5,-6, unpublished (1990).
\bibitem{WV} H.M. Wiseman and J.A. Vaccaro, Phys. Lett.
 {\bf 250 A}, 241 (1998); H.M. Wiseman and Z. Brady,
 Phys. Rev. A {\bf 62}, 023805 (2000).
\bibitem{Zeh99} H.D. Zeh, Electronic report {\tt quant-ph/9908084}.

\end{thebibliography}
\end{document}